\pdfoutput=1
\documentclass[prd,aps,nofootinbib,twocolumn,superscriptaddress,
preprintnumbers,balancelastpage,longbibliography]{revtex4-1}
\usepackage{ae,aecompl}
\usepackage{graphicx}	
\usepackage{amsmath}	
\usepackage{graphicx}
\usepackage{dcolumn}
\usepackage{bm}
\usepackage{graphics}
\usepackage{afterpage}
\usepackage{float}
\usepackage{subfigure}
\usepackage{rotating}
\usepackage{multirow}
\usepackage{tabularx}
\usepackage{booktabs}
\usepackage{multirow}
\usepackage{fancyhdr}
\usepackage{hyperref}
\usepackage[utf8]{inputenc}
\usepackage{theorem}
\usepackage{moreverb}
\usepackage{euscript}
\usepackage{psfrag}
\usepackage{slashed}
\usepackage{mathtools}
\usepackage{makecell}
\usepackage[flushleft]{threeparttable}
\usepackage[english]{babel}
\usepackage{bm}
\usepackage[mathlines]{lineno}
\usepackage{color}
\definecolor{rossoCP3}{cmyk}{0,.88,.77,.40}
\definecolor{darkred}{rgb}{0.6,0,0}
\definecolor{drkgrn}{RGB}{0, 51, 0}
\hypersetup{
     colorlinks   = true,
     citecolor    = blue,
     urlcolor     = magenta,
      linkcolor    = darkred
}

\usepackage{listings}
\usepackage{color,xcolor}

\newcommand{\be}{\begin{equation}}

\newcommand{\ee}{\end{equation}}
\newcommand{\bea}{\begin{eqnarray}}
\newcommand{\eea}{\end{eqnarray}}

\newcolumntype{C}[1]{>{\centering\let\newline\\\arraybackslash\hspace{0pt}}m{#1}}

\lstdefinestyle{python}{
  belowcaptionskip=1\baselineskip,
  breaklines=true,
  frame=L,
  xleftmargin=\parindent,
  language=Python,
  showstringspaces=false,
  basicstyle=\small\ttfamily,
  morekeywords={models, lambda, forms,True,False,None},
  keywordstyle=\bfseries\color{deepgreen!40!black},
  commentstyle=\itshape\color{gray},
  identifierstyle=\color{black},
  stringstyle=\color{deepred},
  rulecolor=\color{gray},
}

\begin{document}

\title{Nuclear Fusion Catalyzed by Doubly Charged Scalars:\\ Implications 
for Energy Production }

\author{Evgeny Akhmedov}
\thanks{{\scriptsize Email}: \href{mailto:akhmedov@mpi-hd.mpg.de}
{akhmedov@mpi-hd.mpg.de}}
\affiliation{Max-Planck-Institut f\"{u}r Kernphysik, Saupfercheckweg 1, 69117 
Heidelberg, Germany}

\date{\today}

\newcommand{\mk}[1]{{\bf #1}}
\newcommand{\om}[1]{\textcolor{red}{#1}}
\newcommand{\sh}[1]{\textcolor{blue}{#1}}

\begin{abstract}
A number of popular extensions of the Standard Model of particle physics 
predict the existence of doubly charged scalar particles $X^{\pm\pm}$. Such 
particles may be long-lived or even stable. If exist, $X^{--}$ could form 
atomic bound states with light nuclei and catalyze their fusion 
by essentially eliminating the Coulomb barrier between them. Such an 
$X$-catalyzed fusion ($X$CF) process does not require high temperatures or 
pressure and may have important applications for energy production. A similar 
process of muon-catalyzed fusion ($\mu$CF) has been shown not to be a viable 
source of energy because of the sticking of negative muons to helium nuclei 
produced in the fusion of hydrogen isotopes, which stops the catalytic 
process. We analyze $X$CF in deuterium environments and show that the 
$X$-particles can only stick to $^6$Li nuclei, which are produced in the 
third-stage reactions downstream in the catalytic cycle. The corresponding 
sticking probability is very low, and, before getting bound to $^6$Li,   
each $X$-particle  can catalyze $\sim 3.5\cdot 10^{9}$ fusion cycles, 
producing $\sim 7\cdot 10^{4}$ TeV of energy. We also discuss the ways of 
reactivating the $X$-particles from the Coulomb-bound (${\rm ^6Li}X$) states, 
which would allow reusing them in $X$CF reactions. 
\end{abstract}

\maketitle

\section{\label{sec:intro}Introduction}
A number of popular extensions of the Standard 
Model (SM) of particle physics predict the existence of doubly charged scalar 
particles $X^{\pm\pm}$. These include the Type-II seesaw \cite{Schechter:1980gr,
Magg:1980ut,Cheng:1980qt,Lazarides:1980nt,Mohapatra:1980yp,Lindner:2016bgg} and 
the Zee-Babu \cite{Zee:1980ai,Babu:1988ki} models of neutrino masses, the 
Left–Right model \cite{Pati:1974yy,Mohapatra:1974hk,Senjanovic:1975rk}, 
the Georgi–Machacek model \cite{Georgi:1985nv,Chanowitz:1985ug,Gunion:1989ci,
Gunion:1990dt,Ismail:2020zoz}, the 3-3-1 model \cite{CiezaMontalvo:2006zt,
Alves:2011kc} and the little Higgs model \cite{ArkaniHamed:2002qx}). 
Doubly charged scalars appear also in simplified models, in which one merely 
adds such scalars in a gauge invariant way in various representations of the 
SM gauge group $SU(2)_L$ to the particle content of the SM. 
The Lagrangian of the model is then complemented by 
gauge-invariant interaction terms involving these new fields 
\cite{Delgado:2011iz,Alloul:2013raa}. 

Doubly charged scalars may be long-lived or even stable 
\cite{Alloul:2013raa,Alimena:2019zri,Acharya:2020uwc,Hirsch:2021wge}. 
As the simplest example, one can add to the SM an uncolored 
$SU(2)_L$-singlet scalar field $X$ with hypercharge 
$Y=2$ \cite{Alloul:2013raa}. The corresponding doubly charged particles will 
couple to the neutral gauge bosons $\gamma$ and $Z^0$ and may also 
interact with the SM Higgs boson $H$ through the  $(H^\dag H)(X^\dag X)$ 
term in the Higgs potential. Gauge invariance allows, in addition, the Yukawa 
coupling of $X$ to right-handed charged leptons, $h_X l_R l_R X+h.c.$
This is the only coupling that makes the $X$-particles unstable 
in this model;  they will be long-lived if the Yukawa coupling constants 
$h_X$ are small. The Yukawa coupling of $X$  may be forbidden by e.g.\ $Z_2$ 
symmetry $X\to -X$, in which case the $X$-scalars will be stable. 

Doubly charged scalar particles are being actively searched for
experimentally, but up to now have not been discovered. 
For discussions of current experimental constraints on the doubly charged 
particles and of the sensitivities to them of future experiments 
see \cite{Alloul:2013raa,Alimena:2019zri,Fuks:2019clu,Padhan:2019jlc,
Acharya:2020uwc,Hirsch:2021wge,Dev:2021axj} and references therein.

In addition to interesting particle-physics phenomenology, doubly charged 
scalars may have important implications for cosmology. In this paper we 
will, however, consider another aspect of their possible existence. 
As we shall demonstrate, doubly charged particles can catalyze fusion of 
light nuclei, with potentially important applications for energy production. 
The negatively charged $X^{--}$ 
(which we will hereafter simply refer to as  
$X$) can form atomic bound systems with the nuclei of light elements, such as 
deuterium, tritium or helium. 
One example is the antihelium-like $(ddX)$ atom 
with the $X$-particle as the ``nucleus'' and two deuterons in the 1$s$ atomic 
state instead of two positrons. (Here and below we use the brackets to denote 
states bound by the Coulomb force). As $X$ is expected to be very heavy, the 
size of such an atomic system will in fact be determined by the deuteron 
mass $m_d$ and will be of the order of the Bohr radius of the $(dX)$ 
ion, $a_d\simeq 7.2$ fm. Similar small-size atomic systems $(N\!N'X)$ can 
exist for other light nuclei $N$ and $N'$ with charges $Z\le 2$. 

Atomic binding of two nuclei to an $X$-particle brings them so close together 
that this essentially eliminates the necessity for them to overcome the 
Coulomb barrier in order to undergo fusion. The exothermic fusion reactions 
can then occur unhindered and do not require high temperature or pressure. 
The $X$-particle is not consumed in this process and can then facilitate  
further nuclear fusion reactions, acting thus as a catalyst. 

This $X$-catalyzed fusion mechanism is to some extent similar to muon 
catalyzed fusion ($\mu$CF) (\cite{Frank,Sakharov,Zeldovich1,Alvarez:1957un,
Jackson:1957zza,Zeldovich2,Zeldovich3,bogd}, see \cite{Zeldovich4,GerPom,
bracci,Breunlich:1989vg,Ponomarev:1990pn,Zeldovich:1991pbl,bogd2} for 
reviews), in which the role of the catalyst is played by singly negatively 
charged muons. $\mu$CF of hydrogen isotopes was once considered a prospective 
candidate for cold fusion. However, already rather early in its studies it 
became clear that $\mu$CF suffers from a serious shortcoming that 
may prevent it from being a viable mechanism of energy production. In the 
fusion processes, isotopes of helium are produced, 
and there is a chance that they will capture on their atomic orbits 
the negative muons present in the final state of the fusion reactions.    
Once this happens, muonic ions ($^3$He$\mu)$ or ($^4$He$\mu)$ are formed, 
which, being positively charged, cannot catalyze further fusion reactions. 
This effect is cumulative; the sticking to helium nuclei thus eventually 
knocks the muons out from the catalytic process, i.e.\ the catalytic 
poisoning occurs. 

Out of all $\mu$CF reactions, the $d-t$ fusion has the smallest muon sticking 
probability, $\omega_s\simeq 10^{-2}$. This means that a single muon will 
catalyze $\sim$100 fusion reactions before it gets removed from the catalytic 
process. The corresponding total produced energy is $\sim$1.7 GeV, which is 
at least a factor of five smaller than the energy 
needed to produce and handle one muon \cite{Jackson:1957zza}. 
In addition, muon's short lifetime makes it impractical to try to dissolve 
the produced ($^3$He$\mu)$ or ($^4$He$\mu)$ bound states by irradiating them 
with particle beams in order to reuse the released muons.
These considerations have essentially killed the idea of using $\mu$CF 
for energy production. 

There were discussions in the literature of 
the possibility of energy generation through the catalysis of 
nuclear fusion by hypothetic heavy long-lived or stable singly 
charged \cite{Zeldovich3,Rafelski:1989pz,
Ioffe:1979tv,Hamaguchi:2006vp} or fractionally charged 
\cite{Zweig:1978sb} particles. However, it has been shown in 
\cite{Zeldovich3,Ioffe:1979tv,Hamaguchi:2006vp} that these processes  
suffer from the same problem of catalytic poisoning 
as $\mu$CF, and therefore they cannot be useful sources of energy. 
In particular, in ref.~\cite{Ioffe:1979tv} it was demonstrated that 
reactivation of the catalyst particles by irradiating their atomic bound 
states with helium nuclei by neutron beams, as suggested in 
\cite{Zweig:1978sb}, would require beams that are about nine orders of 
magnitude higher than those currently produced by most powerful nuclear 
reactors. 

In this paper we consider the fusion of light nuclei catalyzed by doubly 
negatively charged $X$-particles and demonstrate that, unlike $\mu$CF, this 
process may be a viable source of energy. We analyze $X$-catalyzed fusion 
($X$CF) in deuterium environments and show that the  catalytic poisoning may 
only occur in this case due to the sticking of $X$-particles 
to $^6$Li nuclei, which are produced in the fusion reactions of the third 
stage. The corresponding sticking probability is shown to be very low, 
and, before getting bound to $^6$Li, each $X$-particle can catalyze 
$\sim 3.5\cdot 10^{9}$ fusion cycles, producing $\sim 7\cdot 10^{4}$ TeV 
of energy. 
To the best of the present author's knowledge, nuclear fusion 
catalyzed by doubly charged particles has never been considered before.

\section{\label{sec:xcg}$X$-catalyzed fusion 
in deuterium}

We will be assuming that $X$-particles interact only electromagnetically, 
which in any case should be a very good approximation at low energies relevant 
to  nuclear fusion. Let $X$-particles be injected in pressurized D$_2$ gas or 
liquid deuterium. Being very heavy and negatively charged, 
the $X$-particles can easily penetrate D$_2$ molecules and D atoms, 
dissociating the former and ionizing the latter and losing energy 
on the way.  Once the velocity of an $X$-particle becomes comparable to 
atomic velocities ($v\simeq 2e^2/\hbar\sim 10^{-2}c$), it captures a deuteron 
on a highly excited atomic level of the ($dX$) system, which then very 
quickly  de-excites to its ground state, mostly through electric dipole 
radiation and inelastic scattering on the neighboring deuterium atoms. 
As the ($dX$) ion is negatively charged, it swiftly picks up another 
deuteron to form the ($ddX$) atom. The characteristic time of this 
atomic phase of the $X$CF process is dominated by the $X$ moderation time  
and is $\sim 10^{-10}$\,s at liquid hydrogen density 
$N_0=4.25\times 10^{22}$ nuclei/cm$^3$ and $T\simeq 20$K and 
about $10^{-7}$\,s in deuterium gas at $0^\circ$C and pressure of one bar 
(see Appendix~\ref{sec:Xatom}).  
 
After the $(ddX)$ atom has been formed, the deuterons 
undergo nuclear fusion through several channels, see below.   
Simple estimates show that the fusion rates are many orders of magnitudes 
faster than the rates of the atomic formation processes. 
That is, once ($ddX$) [or similar ($N\!N'X$)] atoms are formed, the 
fusion occurs practically instantaneously. The time scale of $X$CF is 
therefore determined by the atomic formation times. 
The rates of the fusion reactions, however,  determine the branching ratios of 
various fusion channels, which are important for the kinetics of the 
catalytic cycle. 

At the first stage of $X$CF in deuterium two deuterons fuse to produce 
$^3$He, $^3$H or $^4$He. In each case there is at least one channel 
in which the final-state $X$ forms an atomic bound state with one of the 
produced nuclei. Stage I fusion reactions are     
\begin{align}
&(ddX)\to {^3\rm He}+n+X &(Q=2.98~{\rm MeV},~29.1\%) 
\tag{1a} \label{eq:r1a}\\
&(ddX)\to ({\rm ^3He}X)+n &(Q=3.89~{\rm MeV},
~19.4\%) \tag{1b} \label{eq:r1b}
\end{align}
\vglue-8mm
\begin{align}
&(ddX)\to {\rm ^3H}+p+X &&(Q=3.74~{\rm MeV},~34.4\%)
\tag{2a} \label{eq:r2a}\\
&(ddX)\to ({\rm ^3H}X)+p &&(Q=4.01~{\rm MeV}, ~6.2\%)
\tag{2b} \label{eq:r2b}\\
&(ddX)\to {\rm ^3H}+(pX) &&(Q=3.84~{\rm MeV},
~0.5\%)
\tag{2c} \label{eq:r2c}
\end{align}
\vglue-8mm
\begin{align}
&(ddX)\to {\rm ^4He}+\gamma+X &&(Q=23.6~{\rm MeV}, ~4\!\cdot\!10^{-9})
\tag{3a} \label{eq:r3a}\\
&(ddX)\to ({\rm ^4He}X)+\gamma &&(Q=24.7~{\rm MeV}, ~3\!\cdot\! 10^{-8})
\tag{3b} \label{eq:r3b}\\
&(ddX)\to {\rm ^4He}+X &&(Q=23.6~{\rm MeV}, ~10.4\%)
\tag{3c} \label{eq:r3c}
\end{align}
Here in the parentheses the $Q$-values and the branching ratios of the 
reactions are shown. In evaluating the $Q$-values we have taken into account 
that the atomic binding of the two deuterons to $X$ in the initial state 
reduces $Q$, whereas the binding to $X$ of one of the 
final-state nuclei increases it. As the Bohr radii of most of the $X$-atomic 
states we consider are either comparable to or smaller than the nuclear 
radii, in calculating the Coulomb binding energies one has to allow for the 
finite nuclear sizes.  We do that by making use of a variational approach, 
as described in Appendix~\ref{sec:Bind}.  

The rates of reactions (\ref{eq:r1b}), (\ref{eq:r2b}), (\ref{eq:r2c}) and 
(\ref{eq:r3b}) with bound $X$-particles in the final states 
are proportional to the corresponding $X$-particle sticking probabilities, 
$\omega_{s}$. The existence of such channels obviously affects the branching 
ratios of the analogous reactions with free $X$ in the final states. 

Radiative reactions (\ref{eq:r3a}) and (\ref{eq:r3b}) have tiny branching 
ratios, which is related to their electromagnetic nature and to the fact that 
for their $X$-less version, $d+d\to{\rm ^4He}+\gamma$, transitions 
of E1 type are strictly forbidden. This comes about because the two fusing 
nuclei are identical, which, in particular, means that they have the same 
charge-to-mass ratio. This reaction therefore proceeds mainly 
through E2 transitions \cite{bogd}. When the deuterons are bound to $X$,  
the strict prohibition of E1 transitions is lifted due to possible 
transitions through intermediate excited atomic states.%
\footnote{\label{fn:1}The author is grateful to M.~Pospelov for raising this 
issue and suggesting an example of a route through which E1 transitions could 
proceed in reaction (\ref{eq:r3b}).}
However, as shown in Appendix~\ref{sec:lift}, 
the resulting E1 transitions are in this case heavily hindered and their 
rates actually fall below the rates of the E2 transitions.  
    
Reaction (\ref{eq:r3c}) is an internal conversion process. 
Note that, unlike for reactions (\ref{eq:r1a}) - (\ref{eq:r3b}), 
the $X$-less version of (\ref{eq:r3c}) does not exist: the process 
$d+d\to ^4$He is forbidden by kinematics. For the details of the calculation 
of the rate of reaction (\ref{eq:r3c}) as well as of the rates of the other 
discussed in this paper reactions, see Appendix~\ref{sec:Sfactor}. 
The relevant $Q$-values of the reactions and sticking probabilities are 
evaluated in Appendices~\ref{sec:Bind} and \ref{sec:sticking}, respectively.

The final states of reactions (\ref{eq:r1a}), (\ref{eq:r2a}), (\ref{eq:r3a}) 
and (\ref{eq:r3c}) contain free $X$-particles which are practically at rest 
and can immediately capture deuterons of the medium, forming again the 
($ddX$) atoms. Thus, they can again catalyze $d-d$ fusion through stage I 
reactions (\ref{eq:r1a})-(\ref{eq:r3c}). 
The same is also true for the $X$-particles in the final state of reaction 
(\ref{eq:r2c}) which  emerge being bound to protons. Collisions of ($pX$) 
with deuterons of the medium lead to fast replacement of the protons 
by deuterons through the exothermic charge exchange reaction 
$(pX)+d\to (dX)+p$ with the energy release $\sim$90~keV 
(see Appendix~\ref{sec:charge}). 
The produced $(dX)$ ion then picks up a deuteron to form the $(ddX)$ atom, 
which can again participate in stage I reactions 
(\ref{eq:r1a})-(\ref{eq:r3c}). 
 
The situation is different for the $X$-particles 
in the final states of reactions (\ref{eq:r1b}) and (\ref{eq:r2b})
forming the bound states with $^3$He and $^3$H, respectively. 
They can no longer directly participate in stage I $d-d$ fusion reactions. 
However, they are not lost for the fusion process: the produced 
(${\rm ^3He}X$) and (${\rm ^3H}X$)  can still pick up deuterons of the medium 
to form the atomic bound states (${\rm ^3He}dX$) and 
(${\rm ^3H}dX$), which can give rise to stage II fusion reactions, which we 
will consider next. 

Before we proceed, a comment is in order. While (${\rm ^3H}X$) is a singly 
negatively charged ion which can obviously pick up a positively charged 
deuteron to form an (${\rm ^3H}dX$) atom, (${\rm ^3He}X$) is a neutral 
$X$-atom. It is not immediately obvious whether it can form a stable bound 
state with $d$, which, if exists, would be a positive ion. In the case of 
the usual atomic systems, analogous (though negatively charged) states 
do exist -- a well-known example is the negative ion of hydrogen H$^-$.
However, the stability of (${\rm ^3He}\,dX$) cannot be directly deduced 
from the stability of H$^-$: in the latter case the two particles 
orbiting the nucleus are identical electrons, whereas for 
(${\rm ^3He}dX$) these are different entities -- nuclei with differing masses 
and charges. Nevertheless, from the results of a general analysis of 
three-body Coulomb systems carried out in \cite{Martin:1998zc,krikeb,armour} 
it follows that the state (${\rm ^3He}dX$) (as well as the bound state 
(${\rm ^4He}dX$) which we will discuss later on) should exist and be stable. 
For additional information see Appendix~\ref{sec:posIons}. 

Once (${\rm ^3He}X$) and (${\rm ^3H}X$), produced in reactions (\ref{eq:r1b}) 
and (\ref{eq:r2b}), have picked up deuterons from the medium and formed the 
atomic bound states (${\rm ^3He}dX$) and (${\rm ^3H}dX$), 
the following stage II fusion reactions occur: 
\begin{align}
\!\!\!\!\!&(^3{\rm He}dX)\to {\rm ^4He}+p+X\!\!\!
&&(Q=17.4~{\rm MeV},\,94\%) 
\!\!\!\tag{4a} \label{eq:r4a}\\
\!\!\!\!\!&({\rm ^3He}dX)\to ({\rm ^4He}X)+p\!\!\!
&&(Q=18.6~{\rm MeV},\,6\%) \nonumber
\!\!\!\tag{4b} \label{eq:r4b}\\
\!\!\!\!\!&({\rm ^3He}dX)\to {\rm ^4He}+(pX)\!\!\!
&&(Q=17.5~{\rm MeV},\,3\!\cdot\!10^{-4})
\!\!\!
\tag{4c} \label{eq:r4c}
\end{align}
\vglue-8.0mm 
\begin{align}
\!\!\!\!\!\!&(^3{\rm H} d X)
\to {\rm ^4He}+n+X 
\!\!\!&&(Q=17.3~{\rm MeV},~96\%)
\tag{5a} \label{eq:r5a}\\
\!\!\!\!\!\!&(^3{\rm H} d X)
\to ({\rm ^4He}X)+n \!\!\!
&&(Q=18.4~{\rm MeV}, ~4\%)
\tag{5b} \label{eq:r5b}
\end{align}
In these reactions vast majority of $X$ bound to $^3$He and $^3$H are 
liberated; the freed $X$-particles can again form $(ddX)$ states and catalyze 
stage I fusion reactions (\ref{eq:r1a})-(\ref{eq:r3c}). The same applies to 
the final-state $X$-particles bound to protons, as was discussed above. 
The remaining relatively small fraction of $X$-particles come out of 
stage II reactions in the form of $({\rm ^4He}X)$ atoms. 
Together with a very small amount of $({\rm ^4He}X)$ produced in 
reaction (\ref{eq:r3b}), they pick up deuterons from the medium and form 
$({\rm ^4He}dX)$ states, which undergo stage III $X$CF reactions:
\begin{align}
&(^4{\rm He} d X)\to {\rm ^6Li}+\gamma+X 
\!\!\!
\!\!\!
&&(Q=0.32~{\rm MeV}, 
~10^{-13})
\tag{6a} \label{eq:r6a}\\
&({\rm ^4He} d X)\to ({\rm ^6Li}X)+\gamma 
\!\!\!
\!\!\!
&&(Q=2.4~{\rm MeV}, ~\,2\!\cdot\! 10^{-8})
\tag{6b} \label{eq:r6b}\\
&({\rm ^4He} d X)\to {\rm ^6Li}+X \!\!\!
\!\!\!\!\!&&(Q=0.32~{\rm MeV}, 
\,\simeq100\%)
\tag{6c} \label{eq:r6c}
\end{align}
In these reactions, almost all previously bound $X$-particles are liberated 
and are free to catalyze again nuclear fusion through $X$CF reactions of 
stages I and II. The remaining tiny fraction of $X$-particles end up being 
bound to the produced ${\rm ^6Li}$ nuclei through reaction (\ref{eq:r6b}). 
However, as small as it is, this fraction is very important for the 
kinetics of $X$CF. The bound states $({\rm ^6Li}X)$ are ions of 
charge +1; they cannot form bound state with positively charged nuclei and 
participate in further $X$CF reaction. That is, with their formation 
catalytic poisoning occurs and the catalytic process stops. 

{}From the branching ratios of stage I, II, and III $X$CF reactions one  
finds that the fraction of the initially injected $X$-particles which 
end up in the $({\rm ^6Li}X)$ bound state is $\sim 2.8\times 10^{-10}$. 
This means that each initial $X$-particle, before getting stuck to a $^6$Li 
nucleus, can catalyze $\sim 3.5\times 10^{9}$ fusion cycles.  

Direct inspection shows that, independently of which sub-channels were 
involved, the net effect of stage I, II and III 
$X$CF reactions is  the conversion of four deuterons to 
a $^6$Li nucleus, a proton and a neutron:
\be
4d\to {\rm ^6Li}+p+n+23.1\,{\rm MeV}\,.
\tag{7} \label{eq:7}
\ee
Therefore, each initial $X$-particle will produce about $7\times 10^4$ TeV of 
energy before it gets knocked out of the catalytic process.  It should be 
stressed that this assumes that the $X$-particles are sufficiently 
long-lived to survive during $3.5\times 10^9$ fusion cycles. From our 
analysis it follows that the slowest processes in the $X$CF cycle are the 
formation of positive ions $({\rm ^3He}dX)$ and $({\rm ^4He}dX)$. 
The corresponding formation times are estimated to be of the order 
of $10^{-8}$\,s (see Appendix~\ref{sec:posIons}). 
Therefore, for the $X$-particles to survive during $3.5\times 10^9$ fusion 
cycles and produce $\sim 7\times 10^4$ TeV of energy, their lifetime $\tau_X$ 
should exceed $\sim 10^2$\,s. For shorter lifetimes the energy produced by a 
single $X$-particle before it gets stuck to a $^6$Li nucleus 
is reduced accordingly. 

\section{\label{sec:acquis} Acquisition and reactivation of $X$-particles}

The amount of energy produced by a single $X$-particle  has to be compared 
with energy expenditures related to its production. 
$X$-particles can be produced in pairs in accelerator experiments, either in 
$l^+l^-$ annihilation at lepton colliders or through the Drell-Yan processes 
at hadronic machines. Although the energy $E\sim 7\times 10^4$ TeV  
produced by one $X$-particle before it gets knocked out of the catalytic 
process is quite large on microscopic scale, it is only about 
10\,mJ. This means that 
$\gtrsim 10^{8}$ $X$-particles are needed to generate 1\,MJ of energy.   
While colliders are better suited for discovery of new particles, 
for production of large numbers of $X$-particles fixed-target accelerator 
experiments are more appropriate. For such experiments the beam energy must 
exceed the mass of the $X$-particle significantly. Currently, plans 
for building such machines are being discussed \cite{Benedikt:2020ejr}.  

The problem is, however, that the $X$-particle production cross section is 
very small. This comes about because of their expected large mass 
($m_X\gtrsim 1$\,TeV/$c^2$) and the fact 
that for their efficient moderation needed to make the formation of $(dX)$ 
atoms possible, $X$-particles should be produced with relatively low velocities.
The cross section $\sigma_p$ of production of $X$-particles 
with mass $m_X\simeq 1$\,TeV/$c^2$ and 
$\beta=v/c\simeq 0.3$ is only $\sim 1$ fb (note that for scalar $X$-particles 
$\sigma_p\propto \beta^3$). As a result, the energy spent on production of an 
$X^{++}X^{--}$ pair will be  by far larger than the energy that can be 
generated by one $X^{--}$ before it gets bound to a $^6$Li nucleus. This 
means that reactivating and reusing the bound $X$-particles multiple times 
would be mandatory in this case. This, in turn, implies that only very 
long lived $X$-particles with $\tau_X\gtrsim 3\times 10^{4}$\,yr 
will be suitable for energy production. 

Reactivation of $X$-particles bound to $^6$Li requires dissociation 
of $({\rm ^6Li}X)$ ions. This could be achieved by irradiating them 
with particle beams, similarly to what was suggested for reactivation of 
lower-charge catalyst particles in ref.~\cite{Zweig:1978sb}.  
However, it would be much more efficient to use instead $({\rm ^6Li}X)$ ions 
as projectiles and irradiate a target with their beam.%
\footnote{We thank M.~Pospelov for this suggestion.} 
The Coulomb binding energy of $X$ to ${\rm ^6Li}$ is about 2 MeV; to strip 
them off by scattering on target nuclei with the average atomic number 
$A\simeq 40$ one would have to accelerate $({\rm ^6Li}X)$ ions to velocities 
$\beta\simeq 0.01$ which, for $m_X\simeq 1$\,TeV/$c^2$, 
corresponds to beam energy $\sim 0.05$ GeV. At these 
energies the cross section of the stripping reaction is $\gtrsim 0.1$\,b, 
and $X$-particles can be liberated with high efficiency in relatively small 
targets. The energy spent on the reactivation of one $X$-particle will then 
only be about $10^{-9}$ of the energy it can produce before sticking to a 
$^6$Li nucleus.

If $X$-particles are stable or practically stable, i.e.\ their lifetime 
$\tau_X$ is comparable to the age of the Universe, there may exist a 
terrestrial population of relic $X$-particles bound to nuclei or (in the 
case of $X^{++}$) to electrons and thus forming exotic nuclei or atoms. The 
possibility of the existence of exotic bound states containing charged 
massive particles was suggested in ref.~\cite{Cahn:1980ss} (see also 
\cite{DeRujula:1989fe}) and has been studied by many authors. 
The concentration of such exotic atoms on the Earth may be very low if 
reheating after inflation occurs at sufficiently low temperatures. 
Note that reheating temperatures as low as a few MeV are consistent 
with observations \cite{Hannestad:2004px}. 
A number of searches for such superheavy exotic isotopes has been 
carried out using a variety of experimental techniques, 
and upper limits on their concentrations were established, see 
\cite{Burdin:2014xma} for a review.

Exotic helium atoms $(X^{++}ee)$ were searched for in the Earth's 
atmosphere using laser spectroscopy technique, and the limit of their 
concentration $10^{-12}-10^{-17}$ per atom over the mass range 20 $-$ 
$10^4$\,GeV/$c^2$ was established~\cite{Mueller:2003ji}. 
In the case of doubly negatively charged $X$, their Coulomb binding to nuclei 
of charge $Z$ would produce superheavy exotic isotopes with nuclear 
properties of the original nuclei but chemical properties of atoms with 
nuclear charge $Z-2$. Such isotopes could have accumulated in 
continental crust and marine sediments. 
Singly positively charged ions 
($^6$Li$X$) and ($^7$Li$X$) chemically behave as superheavy protons; they can 
capture electrons and form anomalously heavy hydrogen atoms. 
Experimental searches for anomalous hydrogen in normal water have put upper 
limits on its concentration at the level of $\sim 10^{-28} - 10^{-29}$ for the 
mass range 12 to 1200 GeV/$c^2$ \cite{smith1} and $\sim 6\times 10^{-15}$ for 
the masses between 10 and $10^5$ TeV/$c^2$ \cite{verkerk}.    

If superheavy isotopes containing relic $X$-particles of cosmological origin 
exist, they can be extracted 
from minerals  e.g.\ by making use of mass spectrometry techniques, and 
their $X$-particles can then be stripped off. To estimate the required 
energy, we conservatively assume that it is twice the energy needed to 
vaporize the matter sample. As an example, it takes about 10 kJ to vaporize 
1\,g of granite \cite{Woskov}; denoting the concentration of $X$-particles in 
granite (number of $X$ per molecule) by $c_X$, we find that the energy 
necessary to extract one $X$-particle is 
$\sim 2.3\times 10^{-18}\,{\rm J}/c_X$. 
Requiring that it does not exceed the energy one $X$-particle can produce 
before getting stuck to a $^6$Li nucleus
leads to the constraint $c_X\gtrsim 2.3\times 10^{-16}$. 
If it is satisfied, extracting  
$X$-particles from granite would allow $X$CF to produce more energy than it 
consumes, even without reactivation and recycling of the $X$-particles. 
Another advantage of the extraction of relic $X$-particles from minerals 
compared with their production at accelerators is that it could work even for 
$X$-particles with mass $m_X\gg 1$\,TeV/$c^2$.  

In assessing the viability of $X$CF as a mechanism of energy generation, 
in addition to pure energy considerations one should 
obviously address  many technical issues 
related to its practical implementation, 
such as collection and moderation of the produced 
$X$-particles and prevention of their binding to the surrounding nuclei 
(or their liberation if such binding occurs), etc. However, the corresponding 
technical difficulties seem to be surmountable~\cite{Goity:1993ih}. 

\section{\label{sec:disc} Discussion} 

There are several obvious ways in which our analysis of $X$CF can be 
generalized. Although we only considered nuclear fusion catalyzed by 
scalar $X$-particles, doubly charged particles of non-zero spin can do 
the job as well. While we studied $X$CF in deuterium, fusion processes 
with participation of other hydrogen isotopes can also be catalyzed by 
$X$-particles.
 
We considered $X$CF taking place in $X$-atomic states.
The catalyzed fusion can also proceed through in-flight reactions occurring 
e.g.\ in $d+(dX)$ collisions. However, because even at the highest attainable 
densities the average distance $\bar{r}$ between deuterons is 
much larger than it is in $(ddX)$ atoms, the rates of in-flight reactions 
are suppressed by a factor of the order of $(\bar{r}/a_d)^3\gtrsim 10^{9}$ 
compared with those of reactions  occurring in $X$-atoms. 

Our results depend sensitively on the properties of positive ions 
$({\rm ^3He} d X)$ and $({\rm ^4He} d X)$, for which we obtained only crude 
estimates. More accurate calculations of these properties and of the formation 
times of these positive ions would be highly desirable. 

The existence of long-lived doubly charged particles may have important 
cosmological consequences. In particular, they may form exotic 
atoms, which have been discussed in connection with the dark matter 
problem \cite{Fargion:2005ep,Belotsky:2006pp,Cudell:2015xiw}. They may also 
affect primordial nucleosynthesis in an important way. 
In ref.~\cite{Pospelov:2006sc} it was suggested 
that singly negatively charged heavy metastable particles may catalyze 
nuclear fusion reactions at the nucleosynthesis era, possibly solving the 
cosmological lithium problem. The issue has been subsequently studied by many 
authors, see refs.~\cite{Pospelov:2010hj,Kusakabe:2017brd} for reviews. 
Doubly charged scalars $X$ may also catalyze nuclear fusion reactions in the 
early Universe and thus may have significant impact on primordial 
nucleosynthesis. On the other hand, cosmology may provide important 
constraints on the $X$CF mechanism discussed here. Therefore, 
a comprehensive study of cosmological implications of the existence of 
$X^{\pm\pm}$ particles would be of great interest. 

To conclude, we have demonstrated that long-lived or stable doubly 
negatively charged scalar particles $X$, if exist, can catalyze nuclear fusion 
and provide a viable source of energy. Our study gives a strong  additional 
motivation for continuing and extending the experimental searches for such 
particles.

{\it Note added.} Recently, the ATLAS Collaboration has reported a 
3.6$\sigma$ (3.3$\sigma$) local (global) excess of events with large 
specific ionization energy loss $|dE/dx|$ in their search for long-lived 
charged particles at LHC \cite{ATLAS:2022pib}. In the complete LHC Run 2 
dataset, seven events were found for which the values of $|dE/dx|$ were 
in tension with the time-of-flight velocity measurements, assuming that 
the corresponding particles were of unit charge. It has been shown in 
\cite{Giudice:2022bpq} that this excess could be explained as being due 
to relatively long-lived doubly charged particles. It would be very 
interesting to see if the reported excess will survive with increasing 
statistics of the forthcoming LHC Run 3.

\section{Acknowledgments}
The author is grateful to Manfred Lindner, Alexei Smirnov and Andreas Trautner 
for useful discussions. Special thanks are due to Maxim Pospelov for numerous 
helpful discussions of various aspects of $X$-catalyzed fusion and 
constructive criticism.

\appendix
\section{\label{sec:Xatom}Atomic processes and $X$-atom formation times} 
\subsection{\label{sec:Xform}Formation times of $X$-atomic systems}

\subsubsection{\label{sec:form}Formation of $(dX)$ ions and of $(ddX)$ 
and $({\rm ^3H}dX)$ atoms}  

Moderation of muons in medium and formation of $\mu$-atoms were considered 
in the classic papers \cite{FT,Wi} which stood the test of time (see e.g.\ 
\cite{C}). The moderation time is practically independent of  
the mass of the ionizing particle and is inversely proportional to square of 
its charge; this allows one to deduce the moderation times for $X$-particles 
by a simple scaling of the muonic case. 
{}From the results of ref.~\cite{Wi} we find that the moderation time of 
$X$-particles from $\beta\equiv v/c\simeq 0.1$ to atomic velocities 
$v\simeq 2e^2/\hbar\simeq 1.5\times 10^{-2}c$ is 
\be
\tau \simeq 6\times 10^{-11}\,{\rm s}\, 
\label{eq:tau}
\ee
at liquid hydrogen density $N_0=4.25\times 10^{22}$ nuclei/cm$^3$ and 
$T\simeq 20$K. It is about $4.8\times 10^{-8}$\,s in deuterium gas at 
$0^\circ$C and pressure of one bar.   

Once an $X$-particle has slowed down to atomic velocities, it gets captured on 
a highly excited state of the ($dX$)-ion, which then de-excites through a 
combination of $\gamma$-ray cascade emission (mostly E1 transitions) and 
inelastic scattering on the neighboring deuterium atoms with their Auger 
ionization. This is similar to de-excitation of highly excited $(\mu d)$ 
atoms in the case of $\mu$CF. In the latter case, the two de-excitations 
processes are generically of comparable rates; at liquid hydrogen density the 
Auger process slightly dominates. The de-excitation to the $(\mu d)$ ground 
state occurs within $t\sim 10^{-12}$\,s \cite{markushin1}. 

In the case of $(dX)$ de-excitation, the radiative processes get enhanced. 
Indeed, the rates of E1 emission are proportional to cube of the energy 
$E_\gamma$ of the emitted $\gamma$ and square of the transition matrix element 
of the electric dipole operator $d_{fi}$. It is easy to see that $E_\gamma$ 
scales linearly with the mass of the atomic orbiting particle $m$, 
whereas $d_{fi}$ scales linearly with the Bohr radius of the system, 
i.e.\ is inversely proportional to $m$. Therefore, the rates of E1 transitions 
scale linearly with $m$. As a result, we find that the rate of radiative 
de-excitation of the ($Xd$) system is larger than that of $(d\mu)$ atom by a 
factor $m_d/m_\mu\sim 20$. One can therefore expect the de-excitation of  
$(dX)$ ions to be at least as fast as that for the muonic deuterium, i.e.\ to 
occur within $\sim 10^{-12}$\,s. 

The produced negative ion $(dX)$ can pick up another deuteron from the medium 
to form a highly excited state of the $(ddX)$ atom, which can de-excite 
through the same processes as $(dX)$. In addition, being electrically neutral, 
$(ddX)$ can penetrate deep inside the neigbouring deuterium atoms and 
experience the electric field of their nuclei. This leads to Stark mixing 
effects which further accelerate the de-excitation processes 
\cite{betheLeon,markushin1}.  

The situation is quite similar for the formation time of $({\rm ^3H}dX)$ 
atoms. An $({\rm ^3H}X)$ negative ion produced in reaction (\ref{eq:r2b}) 
picks up a deuteron from the medium to form a highly excited state 
of $({\rm ^3H}dX)$ atom. The latter then de-excites as described above 
for the $(ddX)$ atom, within approximately the same time interval.

\subsubsection{\label{sec:charge}Charge exchange reaction $d+(pX)\to(dX)+p$}

A simple and accurate estimate of the cross section of the muon exchange 
reaction $d+(p\mu)\to(d\mu)+p$, 
based on dimensional analysis and the fact that low-energy cross sections 
of inelastic processes are inversely proportional to the relative velocities 
of the colliding particles, was given in \cite{Zeldovich4}. 
The cross section of the reaction $d+(pX)\to(dX)+p$ can be estimated 
similarly, which yields  
\be
\sigma_c\simeq 4\pi a_p^2 f v_*/v\,.
\label{eq:sigmaChExch1}
\ee
Here $a_p=\hbar/(2\alpha m_p c)=1.44\times 10^{-12}$\,cm is the Bohr radius 
of $(pX)$ atom, $v$ and $v_*$ are the relative velocities of the 
involved particles in the initial and final states, respectively, and $f$ is 
a constant of order unity. Taking into account that the relative velocities of 
the initial-state particles are very small and the $Q$-value 
of the reaction $d+(pX)\to(dX)+p$ is $\simeq 90$ keV, we find 
$v_*\simeq 1.4\times 10^{-2}c$, which gives 
\be
\sigma_c v \simeq 10^{-14}
\,{\rm cm^3/s}\,.
\label{eq:sigmaChExch2}
\ee
For the rate $\lambda_c$ and the characteristic time $t_c$ of this 
reaction we then find, at liquid hydrogen density 
$N_0=4.25\times 10^{22}$ nuclei/cm$^3$, 
\be
\lambda_c=\sigma_c v N_0\simeq 
4\times 10^{8}\,{\rm s^{-1}}\,,\quad
t_c=\lambda_c^{-1}\simeq 2.5\times 10^{-9}\,{\rm s}\,. 
\label{eq:lambdaC}
\ee

\subsubsection{\label{sec:posIons} 
Positive ions $({\rm ^3He} d X)$ and 
$({\rm ^4He} d X)$ and time scales of their formation}

The complexes $({\rm ^3He}dX)$ and $({\rm ^4He}dX)$ are positive ions. 
Such systems  can be considered as composed of a tightly bound ``inner core'', 
represented by a neutral (${\rm He}X)$ atom, and a deuteron, weakly bound to 
the core by atomic polarization effects. The neutral atoms in the inner cores 
are slightly perturbed by the presence of an external deuteron 
and are characterized by the binding energies and Bohr radii approximately 
equal to those of the corresponding $({\rm ^3He}X)$ or $({\rm ^4He}X)$ atoms 
in the absence of the additional deuteron. 
The extraneous deuteron is bound on a 1$s'$ orbit characterized by a larger 
radius and much smaller binding energy. 

It is not immediately obvious if such exotic 
atomic systems are actually stable; 
in particular, their stability cannot be deduced from the stability of 
negative ion of hydrogen H$^-$ familiar from the usual atomic physics.   
Stability of three-body Coulomb systems with arbitrary masses and charges 
of the particles was studied in a number of papers, see e.g.\ 
\cite{Martin:1998zc,krikeb,armour}. 
{}From their general results it follows that 
the states (${\rm ^3He}dX$) and (${\rm ^4He}dX$) should actually be stable.  
This can be seen from fig.~8 of ref.~\cite{Martin:1998zc}, fig.~3 of 
\cite{krikeb} or fig.~13 of \cite{armour}, where the stability regions are 
shown in the case of arbitrary fixed masses of the particles and the charge 
$q_1=1$ as a function of $q_2^{-1}$ and $q_3^{-1}$. 
Here $q_1$ is the absolute value of the charge of the particle which is 
opposite to the other two ($q_1=Z_Xe$ in the case we consider), 
whereas $q_2$ and $q_3$ are the charges of the same-sign particles, with the 
convention $q_2\ge q_3$. As the stability depends on the ratios of the 
charges and not on their absolute values, $q_1$ was set equal to unity for 
convenience. With such a normalization, we have 
$q_2=1$, $q_3=1/2$.  It can be seen from the above mentioned figures that 
the point $(q_2^{-1},q_3^{-1})=(1,2)$ is inside the stability region, that is, 
positive ions (${\rm ^3He}dX$) and (${\rm ^4He}dX$) must be stable. 
This point is, however, rather close to the border of the stability region, 
which reflects the relative smallness of the binding energy of the 
deuteron (``deuteron affinity''). 

We have attempted a variational calculation of the binding energies of such 
positive ions using simple two- and three-parameter Hylleraas-type 
trial wave functions which were able to predict the stability of H$^-$ 
ion, but found no binding of deuteron. This is apparently 
related to the fact that (${\rm He}dX$) ions, 
which have nuclei of differing mass and charge on their atomic orbits, are 
more complex than H$^-$ ion, whose two electrons are identical particles. 
A qualitative analysis of the properties of such systems would therefore 
necessitate calculations with more sophisticated trial wave functions. 
This would require a dedicated study, which is beyond the scope of the 
present paper.  

In the absence of an actual calculation, we have to resort 
to semi-quantitative methods. In doing that, we will be using the properties 
of negative ion H$^-$ as a starting point, but 
will also take into account 
the peculiarities of (${\rm ^3He}dX$) and (${\rm ^4He}dX$) ions. In the case 
of H$^-$, the radius of the outer electron's orbit is about a factor 
3.7 larger than that of the inner electron \cite{hoga}, and the binding energy 
of the outer electron 
(electron affinity) is about 18 times smaller than that of the inner one. 
Taking into account tighter binding of the inner core in the case 
of the positive ions we consider, we assume the radii 
$a$ of their external orbits and the deuteron binding energies $E_{bd}$ to be, 
respectively, a factor of $\sim 30$ larger and three orders of magnitude 
smaller than those of the corresponding $({\rm He}X)$ atoms. 
Factor $\sim 30$ increase for $a$ compared with the inner core radius
is obtained as follows: we multiply the factor 3.7 hinted by H$^-$ ions by
$2\times 2=4$ due to the $X$ particle and the He nucleus each having charge 
2 and by the ratio of the mass of $^3$He (or $^4$He) to the deuteron mass. 
This gives factor \mbox{$\sim$\,22 $-$ 30}. For the binding energy of the 
extra deuteron we take into account that it scales as $a^{-2}$. 
We therefore choose 
\begin{align}
&({\rm ^3He}dX): \quad 
a\simeq 7\times 10^{-12}\,{\rm cm}\,,\quad 
E_{bd}\simeq 1.2\,{\rm keV}\,,
\label{eq:Ebd1}
\\
&({\rm ^4He}dX): \quad a\simeq 5\times 10^{-12}\,{\rm cm}\,,
\quad
E_{bd}\simeq 1.6\,{\rm keV}\,.
\label{eq:Ebd2}
\end{align}

The formation of (${\rm ^3He}dX$) and (${\rm ^4He}dX$) 
ions can proceed as follows. An (${\rm ^3He}X$) atom  
produced in reaction (\ref{eq:r1b}) 
collides with the neighboring D$_2$ molecules, dissociating them and picking 
up one of their deuterons through the exothermic reaction 
\be
({\rm ^3He}X)+{\rm D}_2 \to ({\rm ^3He}dX)+d+2e^-\,.
\label{eq:da}
\ee
This is the dissociative attachment (DA) mechanism, 
analogous to the one by which H$^-$ ions are produced in 
$e^-$+H$_2\to$H$^-$+H reactions.  An important difference is, however, that 
what is attached is now a nucleus (deuteron) rather than an electron. The 
formation of (${\rm ^4He}dX$) ions from (${\rm ^4He}X$) atoms 
produced in reaction (\ref{eq:r4b}) and (\ref{eq:r5b}) proceeds similarly.  
[Note that a tiny fraction of (${\rm ^4He}dX$) ions is produced directly 
in stage I reaction (\ref{eq:r3b})].

As the $Q$-values of the formation reactions of (${\rm ^3He}dX$) and 
(${\rm ^4He}dX$) ions are about two orders 
of magnitude larger than the dissociation energy of D$_2$ molecules and 
the ionization potential of D atoms, these processes are actually similar to 
the usual charge exchange reactions on free particles, except that most of 
the released energy is now carried away by the final-state electrons. 
The rates and characteristic times of these processes can therefore be 
estimated using the expressions similar to those in 
eqs.~(\ref{eq:sigmaChExch2}) and (\ref{eq:lambdaC}). 
This gives, at the liquid hydrogen density,  
\be
\lambda_{\rm DA}\sim 5\times 10^7\,{\rm s}^{-1}\,,\quad 
t_{\rm DA}=\lambda_{\rm DA}^{-1}\sim 2\times 10^{-8}\,{\rm s}\,. 
\label{eq:tDA}
\ee
\subsection{\label{sec:Bind}Atomic binding energies of light nuclei in 
$X$-atoms
and $Q$-values of fusion reactions
}
The $Q$-value of an $X$CF reaction can be found by subtracting 
from the $Q$-value of the corresponding $X$-less reaction 
the atomic binding  energy of the nuclei in the initial state and adding to 
it the binding energy of one of the produced nuclei to $X$ in the final 
state, when such bound states are formed. We therefore first find the 
relevant binding energies. 

\subsubsection{\label{sec:antiHe}Antihelium-like $(ddX)$ atom} 
The total binding energy of helium atom is $79.005$ eV. 
{}From this value, the binding energy of ($ddX$) atom is obtained to a very 
good accuracy by the simple rescaling with the factor 
$m_d/m_e$, which gives $E_b(ddX)=0.290$ MeV.

\subsubsection{\label{sec:other}Other $X$-atoms with Coulomb-bound light 
nuclei} 

For atomic $(N N' X)$ states other than $(ddX)$ we first consider 
hydrogen-like atoms $(N X)$ and than estimate the atomic binding energy of 
the additional nucleus $N'$ ($m>m'$ is assumed). In the limit of pointlike 
nuclei the ground-state wave function $\psi_{1s}(r)$, the Bohr radius $a$ and 
the binding energy $E_b^0$ of an $(NX)$ state are 
\begin{widetext}
\be
\psi_{1s}(r)=\frac{1}{\sqrt{\pi a^3}}e^{-r/a}\,,\qquad
a=\frac{\hbar^2}{
Z_X Z e^2 m}=\frac{1}{Z_X Z\alpha}
\frac{\hbar }{m c}\,,\qquad 
E_b^0=|E_{1s}^0|=
\frac{1}{2}(Z_X Z \alpha)^2 m c^2\,.
\label{eq:h-like}
\ee 
\end{widetext}
Here $Ze$ and $m$ are the charge and the mass of the nucleus $N$, and $-Z_Xe$ 
is the charge of the $X$-particle ($Z_X=2$ in the case under discussion). 

For most of the nuclei we consider, Bohr radii of the $(NX)$ atomic 
states are either comparable to or smaller than the nuclear radii,  
and the approximation of pointlike nuclei is rather poor. We therefore allow 
for finite nuclear sizes  by making use of a variational approach. 
We consider nuclei as uniformly charged balls of radius $R$  
and employ the simple one-parameter test wave function of Fl\"{u}gge and 
Zickendrant, which has the correct asymptotics for both large and small $r$ 
\cite{fluegge1,fluegge2}:
\be
\psi(r) = 
N(\lambda) \left(1+\frac{\lambda r}{2R}\right) e^{-\frac{\lambda r}{2R}}\,,
\quad 
N(\lambda)=
\sqrt{\frac{1}{7\pi}\left(\frac{\lambda}{2R}\right)^3}\,.
\label{eq:fl1}
\ee
Here $\lambda$ is the variational parameter. 
With this wave function, the expectation value of the energy of the system is 
\begin{widetext}
\be
E(\lambda)=\frac{3}{56}\frac{\hbar^2}{m R^2}\left\{\lambda^2+\frac{R}{a}
\left[\frac{216}{\lambda^2}-28-e^{-\lambda}\left(\frac{216}{\lambda^2}
+\frac{216}{\lambda}+80+14\lambda+\lambda^2\right)\right]\right\}.
\label{eq:fl2}
\ee
\end{widetext}
We minimize it numerically,  
which yields the ground-state energy $E_{1s}$, the binding energy of the 
system being $E_b(R)=|E_{1s}|$. 
The corresponding value of 
$\lambda$ determines, through eq.~(\ref{eq:fl1}), 
the ground-state wave function of the system. It will be used in the 
calculations of the sticking probabilities in sec.~\ref{sec:stick1} below. 

We calculate the binding energies of $(NX)$ states for two different choices 
of the values of the nuclear radii $R$. First, we employ the frequently 
used expression $R=R_N\equiv 1.2A^{1/3}$\, fm, where $A$ is the atomic number 
of the nucleus $N$. Second, we make use of the experimentally measured rms 
charge radii $r_{Nc}\equiv \langle r_c^2\rangle_N^{1/2}$~\cite{Angeli:2013epw} 
and set the nuclear radii equal to $R_{Nc}\equiv (5/3)^{1/2}r_{Nc}$, 
which is the relation between $R_{Nc}$ and $r_{Nc}$ in 
the uniformly charged ball model of the nucleus. 
The results are presented in Table~\ref{tab:bind1} along with the Bohr radii 
$a$ and the binding energies for pointlike nuclei $E_b^0$. Note that for our 
further calculations we use the results based on the experimentally measured 
nuclear charge radii, which are presumably more accurate. 

As a test, we also performed similar calculations for atomic systems $(NC)$ 
with $C$ a singly negatively charged heavy particle, for which the binding 
energies were previously found in~\cite{Pospelov:2006sc}. Our results are in 
good agreement with those of ref.~\cite{Pospelov:2006sc}, 
the difference typically being within 1\%. 

The binding energies of the $X$-atoms in the initial states of the 
$X$CF reactions other than $(ddX)$ are found as follows. 
For $({\rm ^3H}dX)$ atoms, we add to the binding energy of negative 
$({\rm ^3H}X)$ ion the deuteron binding energy found 
through the variational procedure described above, assuming deuteron to 
be a pointlike particle in the Coulomb field of a nucleus of charge 
$Z_{\rm ^3H}-Z_X=-1$ and radius equal to that of $^3$H nucleus. 
For positive ions $({\rm ^3He}dX)$ and $({\rm ^4He}dX)$, we add to the binding 
energies of $({\rm ^3He}X)$ and $({\rm ^4He}X)$ atoms the deuteron binding 
energy $E_{bd}$ given in eqs.~(\ref{eq:Ebd1}) and (\ref{eq:Ebd2}), 
respectively. The obtained binding energies are then used for calculating the 
$Q$-values of the $X$CF reactions under consideration. The results are shown 
in the third column of Table~\ref{tab:general}.

\begin{table*}[t]
\centering
\begin{tabular}{|c|c|c|c|c|c|c|c|}
\hline
\hline
Bound state & Bohr radius $a$ (fm) & $r_{Nc}$ (fm) \cite{Angeli:2013epw} 
&$R_N=1.2A^{1/3}$ (fm) 
& $R_{Nc}$ (fm) & $E_b(R_N)$ (MeV) 
& $E_b(R_{Nc})$ (MeV) & $E_b^0$
(MeV) \\[0.2em]
\hline
($pX$)  & 14.4 & 0.8783 & 1.20 & 1.134 & 0.096 & 0.096  & 0.100     \\
($dX$)  & 7.20 & 2.142 & 1.51  
& 2.765 & 0.189 & 0.183  & 0.200     \\
(${\rm^3H}X$)  & 4.81 & 1.759 & 1.73  
& 2.271 &
0.276 & 0.268 & 0.299    \\
(${\rm ^3He}X$) & 2.41 & 1.966 & 1.73   
&2.538 & 1.00 & 0.905  & 1.196  \\
(${\rm ^4He}X$) & 1.81 & 1.676 & 1.905  
& 2.163 & 1.202 & 1.153  & 1.588 \\
(${\rm ^6Li}X$)  & 0.805 & 2.589 & 2.18  
& 3.342 & 2.680 & 2.069  & 5.369 \\
\hline
\end{tabular}
\caption{Properties of $(NX)$ bound states. Third and fifth columns show 
experimental values of rms charge radii $r_{Nc}\equiv \langle r_c^2
\rangle_N^{1/2}$ from ref.~\cite{Angeli:2013epw} and the corresponding nuclear 
radii found as $R_{Nc}=(5/3)^{1/2}r_{Nc}$. 
$E_b(R_N)$ and $E_b(R_{Nc})$ are 
binding energies calculated for the corresponding values of nuclear radii; 
$E_b^0=(Z_X Z\alpha)^2mc^2/2$ is 
binding energy in the limit of pointlike nuclei. }
\label{tab:bind1}
\end{table*}

\subsection{\label{sec:sticking}Sticking probabilities and related issues}  

\subsubsection{\label{sec:stick1}Sticking probabilities in the sudden 
approximation}

To evaluate the probability $\omega_s$ that the $X$-particle in the 
final state of a fusion process will stick to one of the produced 
nuclear fragments we make use of the fact that nuclear reactions of 
$X$CF occur on time scales that are much shorter than the 
characteristic $X$-atomic time. Indeed, the characteristic time of 
$X$-atomic processes is $t_{\rm at}\sim  a_d/v_{at}
\sim\hbar^3/(4m_d e^4)\simeq 1.6\times 10^{-21}$\,s, 
whereas the fusion reactions of $X$CF occur on the 
nuclear time scales $\lesssim 10^{-23}$\,s. This disparity between the 
atomic and nuclear time scales in $X$CF allows one to use the sudden 
approximation \cite{Migdal} for evaluating the $X$-sticking probabilities 
$\omega_s$. 

Consider the reaction $(N_1N_2X)\to N_3+N_4+X$. Just before the fusion 
occurs, the nuclei in the $(N_1N_2X)$ atom approach each other to a distance 
of the order of the range of nuclear force. The atomic wave function  
therefore adiabatically goes over into that of the hydrogen-like atom with 
a nucleus of mass $m_i=m_1+m_2$ and charge $Z_i=Z_1+Z_2$ orbiting the $X$ 
particle. We denote this wave function $\psi_i$. As the fusion occurs suddenly 
(compared with the atomic time scale), the state of the atomic system 
immediately after the fusion will be described by the same wave function. 
Transition amplitudes can then be found by projecting it 
onto the proper final states. 
Let the velocity of the produced nucleus $N_3$ of mass $m_3$ be $\vec{v}$. 
As the final-state $X$-particle is practically at rest, this is also the 
relative velocity of $N_3$ and $X$.  
The probability that $N_3$ will get captured by $X$ and form a 
bound state with it is then 
\be
\omega_s=\sum_\alpha\Big|\int \psi_{f\alpha}^*\psi_i 
e^{-i\vec{q}\vec{r}}dV\Big|^2\,,
\label{eq:omega1}
\ee 
where $\psi_{f\alpha}$ is the wave function of the final $(N_3X)$ state and  
$\vec{q}=m_*\vec{v}/\hbar$, 
$m_*=m_3 m_X/(m_3+m_X)\simeq m_3$ being the reduced mass of 
the $N_3 X$ system. The sum in (\ref{eq:omega1}) is over all the 
bound states of hydrogen-like atom $(N_3X)$. 

The case of radiative fusion reactions $(N_1N_2X)\to 
N_3+\gamma+X$ is considered quite similarly. 
However, due to different kinematics, the values of $q=|\vec{q}|$ are related 
differently to the $Q$-values for these reactions. For the non-radiative 
reactions we have $q=\sqrt{2\mu Q}/\hbar$ with $\mu=m_3 m_4/(m_3+m_4)$, 
whereas for the radiative 
ones $q\simeq Q/(\hbar c)$.

The main contribution to $\omega_s$ comes from the transition to the ground 
state of $(N_3X)$ atom, with total contribution of all the excited states 
being less than 20\% \cite{Zeldovich3}. For our estimates we shall therefore 
restrict ourselves to transitions to the ground states. 
The functions $\psi_i$ and $\psi_f$ are then 
the wave functions of the 1$s$ states of the hydrogen-like 
atoms with masses and charges of the atomic particles $m_i=m_1+m_2$, 
$Z_i=Z_1+Z_2$ and  $m_f=m_3$, $Z_f=Z_3$, respectively. 

To take into account the finite size of the nuclei, we use the wave 
functions (\ref{eq:fl1}) with the substitutions $\lambda\to 
\lambda_{i,f}$, where the variational parameters $\lambda_{i,f}$ are found 
from the minimization of $E(\lambda)$ defined in (\ref{eq:fl2}) with the 
replacements $a\to a_{i,f}$ and $R\to R_{i,f}$. The Bohr radii $a_{i,f}$ are 
given by the standard formula (see eq.~(\ref{eq:Bohr}) below); 
the nuclear radii $R_{i,f}$ can be found from the rms nuclear charge radii as 
discussed in sec.~\ref{sec:other}. To find $R_f$, we 
can directly use  the experimentally measured rms charge radius of the 
nucleus $N_3$. For the initial state, we 
approximate the rms charge radius $r_i$ of the compound nucleus $N_1N_2$ as 
\be
r_i\simeq (r_{N{_1} c}^3+r_{N_{2} c}^3)^{1/3}\,,
\label{eq:liqdr1}
\ee  
where $r_{N_{1}c}\equiv\langle r_c^2\rangle_{N_1}^{1/2}$ and 
$r_{N_{2}c}\equiv\langle r_c^2\rangle_{N_2}^{1/2}$ are the experimentally 
measured rms charge radii of the $N_1$ and $N_2$ nuclei, respectively. 
Note that eq.~(\ref{eq:liqdr1}) corresponds to the liquid drop model of  
nucleus. Eq.~(\ref{eq:omega1}) then gives for the sticking probability 
\be
\omega_s=\left[\frac{32\pi N_i N_f\,\kappa}{(\kappa^2+q^2)^3}
\bigg(\kappa^2+\frac{3\kappa_i\kappa_f(\kappa^2-q^2)}{\kappa^2+q^2}\bigg)
\right]^2,
\label{eq:omega_s1}
\vspace*{-2mm}
\ee
where 
\be
N_{i,f}=N(\lambda_{i,f})\,,\quad \kappa_{i,f}=
\frac{\lambda_{i,f}}{2R_{i,f}}\,,\quad \kappa=\kappa_i+\kappa_f\,.
\label{eq:Nif}
\ee

For comparison, we also calculate the sticking probabilities $\omega_{s0}$ 
neglecting the nuclear size, i.e.\ employing the usual ground-state 
wave functions of the hydrogen-like atoms with pointlike nuclei  
(\ref{eq:h-like}). From eq.~(\ref{eq:omega1}) we find 
\be
\omega_{s0}=\frac{(2a_r)^6}{(a_i a_f)^3}\frac{1}{(1+q^2 a_r^2)^4}\,,
\label{eq:omega_s2}
\ee
where 
\be
a_{i,f}=\frac{\hbar}{Z_X Z_{i,f}\alpha m_{i,f}c}\,,\qquad
a_r=\frac{a_i a_f}{a_i+a_f}\,.
\label{eq:Bohr}
\vspace*{2mm}
\ee
The obtained values of the sticking probabilities $\omega_s$ and 
$\omega_{s0}$ are shown in the fourth and fifth columns of 
Table~\ref{tab:general}. 

\subsubsection{\label{sec:lift}Lifting the prohibition of radiative E1 
transitions for Coulomb-bound nuclei}  

The radiative nuclear fusion reaction $d+d\to {\rm ^4He}+\gamma$ has a tiny 
branching ratio. This is because it proceeds mostly through E2 electromagnetic 
transitions, as E1 transitions are strictly forbidden for fusions of identical 
particles. Indeed, after the separation of the irrelevant 
center-of-mass motion, one finds that for particles of charges $q_{1,2}$ and 
masses $m_{1,2}$ the effective charge of the electric dipole operator is  
$q=(q_1 m_2-q_2 m_1)/(m_1+m_2)$, which vanishes 
when the two particles have the same charge-to-mass ratio. 
 
The situation may be different for $X$CF, when the fusing particles 
are bound to an orbit of an $X$-atom. Fusion may then proceed through the 
transition to an intermediate excited atomic state, which then de-excites 
via atomic electric dipole radiation. 
In this mechanism the motion of the center of mass of the initial-state 
nuclei plays a central role, and the effective charge of the dipole operator 
does not vanish. Consider the $X$CF reaction 
\be
(ddX)~\to~ ({\rm ^4He}X)^*~\to~ ({\rm ^4He}X)+\gamma\,, 
\label{eq:interm}
\ee 
where $(ddX)$ is in its ground $(1s)^2$ state and $({\rm ^4He}X)^*$ is an 
excited state of $({\rm ^4He}X)$ atom which can decay through an E1 
transition. For definiteness, we take $({\rm ^4He}X)^*$ to be a state 
with the principal quantum number $n=3$ (contributions of states with higher 
$n$ are in general suppressed as $1/n^3$).  
Conservation of the total angular momentum and parity in strong interactions 
responsible for the fusion process imply that the intermediate $n=3$ state 
can be either 3$s$ or 3$d$. The excited $({\rm ^4He}X)^*$ state can then decay  
through E1 $\gamma$-emission to the 2$p$ state of $({\rm ^4He}X)$ atom, 
which will eventually de-excite to the ground state of $({\rm ^4He}X)$. Thus, 
the radiative fusion reaction (\ref{eq:r3b}) might proceed through an E1 
transition from an excited state of $({\rm ^4He}X)$ rather than through the 
usual E2 transitions.

It is easy to see, however, that this does not lead to any appreciable 
increase of the rate of reaction (\ref{eq:r3b}). Indeed, the amplitude of the 
process in eq.~(\ref{eq:interm}) contains the product of the amplitude of 
fusion with the formation of  $({\rm ^4He}X)^*$ and the amplitude of its 
subsequent E1 de-excitation. The rate of the process (\ref{eq:interm}) is 
therefore proportional to the probability 
that ${\rm ^4He}$ produced as a result of the fusion reaction is bound to $X$ 
in an excited state of the $({\rm ^4He}X)$ atom rather than being ejected. 
The latter can be found in the sudden approximation by making use of 
the expression on the right-hand side of  
eq.~(\ref{eq:omega1}) with $\psi_{f\alpha}$ 
being the wave functions of the $3s$ and $3d$ states. 
Using for our estimates the hydrogen-like wave functions for pointlike 
nuclei, we find for the corresponding probabilities    
\be
P_{3s}=
\frac{2^8}{3^3}
\frac{(q^2 a^2)^2(q^2 a^2+\frac{16}{27})^2}{(q^2a^2
+\frac{16}{9})^8}\,,
\quad P_{3d}=
\frac{2^{17}}{3^9}\frac{(q^2 a^2)^2}{(q^2a^2+\frac{16}{9})^8}\,. 
\label{eq:}
\ee
Here $q\simeq Q/(\hbar c)$ is the momentum transfer to the produced $^4$He 
nucleus divided by $\hbar$, and $a$ is the Bohr radius of ($^4$He$X$) atom. 
Note that these probabilities are suppressed for both large and small $qa$. 
Large-$qa$ quenching is a result of fast oscillations of the factor 
$e^{-i\vec{q}\vec{r}}$ in the integrand of (\ref{eq:omega1}), whereas 
the suppression at small $qa$ is a consequence of the orthogonality of the 
wave functions $\psi_{f\alpha}$ and $\psi_i$. 
For process (\ref{eq:interm}) we have $qa\simeq 0.22$, which yields  
$P_{3s}\simeq 8\times 10^{-3}$, $P_{3d}\simeq 1.1\times 10^{-2}$. In 
estimating the rate of reaction (\ref{eq:interm})  we also have to take into 
account that the photon emission process is non-resonant. This leads to an 
additional suppression by the factor 
$\sim(\Delta E/E_\gamma)^2\simeq 9\times 10^{-5}$, where 
$\Delta E\simeq 0.2$\,MeV is the energy difference between the $n=3$ and $n=2$ 
states of $({\rm ^4He}X)$ atom and $E_\gamma\simeq Q\simeq 24$\,MeV is the 
energy of the emitted $\gamma$. As a result, the rate of the $X$-atomic E1 
transition gets suppressed by a factor $1.7\times 10^{-6}$ compared with what 
is expected for a typical E1 transition.  

To assess the importance of the $X$-atomic channel (\ref{eq:interm}) 
of reaction (\ref{eq:r3b}) we use the simple estimate of the rates of 
electric multipole transitions from ref.~\cite{weiss}: 
\be
\Gamma({\rm E}l)\simeq \frac{2(l+1)}{l[(2l+1)!!]^2}\left(\frac{3}{l+3}\right)^2
\alpha\, \left(\frac{E_\gamma R}{\hbar c}\right)^{2l}\frac{E_\gamma}{\hbar}\,.
\label{eq:weiss}
\ee
Here $R$ is the nuclear radius $R_{N}$ for nuclear transitions and the size 
of the atomic system $a$ for transitions between atomic states. 
Note that, in the case of $X$-atoms with light nuclei, the nuclear and  
atomic radii are of the same order of magnitude (see 
Table~\ref{tab:bind1}). 
For the process under consideration one could expect, neglecting the 
suppression of E1 transitions, $\Gamma({\rm E1})_{\rm unsup.}
/\Gamma({\rm E2})\sim (625/12)(E_\gamma a/\hbar c)^{-2}
\sim 800$. Taking into account the discussed above suppression factor, 
we arrive at $\Gamma({\rm E1})/\Gamma({\rm E2})\sim 1.4\times 10^{-3}$. 

Thus, although transitions through excited atomic states of $({\rm ^4He}X)$ 
atom lift the forbiddance of electric dipole radiation in the  
fusion reaction (\ref{eq:r3b}), the resulting E1 transition is heavily 
hindered, and its contribution to the rate of the process can be 
neglected.

A very similar argument applies to the radiative fusion reaction 
(\ref{eq:r6b}). Although the fusing nuclei, ${\rm ^4He}$ and $d$, are 
not identical in this case, they have nearly the same 
charge-to-mass ratio. As a result, in the corresponding $X$-less 
reaction E1 transitions are heavily suppressed, and the reaction 
proceeds primarily through E2 radiation. The $X$CF reaction (\ref{eq:r6b}) 
could go through an excited atomic states of $({\rm ^6Li}X)$ atom. 
We find, however, that in this case the suppression of the atomic E1 
transition is even stronger than it is for reaction (\ref{eq:r3b}). Indeed, 
the $Q$-value of the reaction is rather small 
($Q\simeq 2.4$\,MeV), leading to $qa\simeq 10^{-2}$. Assuming again 
transitions through $n=3$ atomic states, we find for the probabilities of 
formation of the excited states $({\rm ^6Li}X)^*$ 
the values $P_{3s}\simeq 3\times 10^{-10}$, $P_{3d}\simeq 6\times 10^{-10}$. 
The suppression factor due to the non-resonant 
nature of the radiative transitions from the $3s$ and $3d$ states to the 
$2p$ state of $({\rm ^6Li}X)$ is  $(\Delta E/E_\gamma)^2\simeq 0.02$.
Altogether, this gives for the atomic E1 transitions in process
(\ref{eq:r6b}) the suppression factor $\sim2\times 10^{-11}$, which makes 
them completely irrelevant.

\section{\label{sec:Sfactor}Astrophysical $S$-factors, reaction factors and 
fusion rates}
In this section we give the details of our calculations of the cross sections, 
rates and branching ratios of the $X$CF reactions under discussion. 
 \vspace{-6mm}

\subsection{\label{sec:cross}Cross sections and reaction factors}

The cross section of a fusion reaction of nuclei $N_1$ and  $N_2$ of 
masses $m_{1,2}$ and charges $Z_{1,2}$ is usually written as \cite{clayton} 

\be
\sigma(E)=\frac{S(E)}{E} e^{-2\pi\eta_{12}}\,,
\label{eq:crsec1}
\ee
where $S(E)$ is the so-called astrophysical factor, $E$ is the c.m.s. energy  
and $\eta_{12}$ is the Sommerfeld parameter:  
\be
\eta_{12}=\frac{Z_1 Z_2 e^2}{\hbar v}=Z_1 Z_2\alpha\sqrt{\frac{\mu c^2}{2E}}
 \,,\qquad 
\mu=\frac{m_1 m_2}{m_1+m_2}\,.
\label {eq:eta1}
\ee
Here $v$ is the relative velocity of the fusing particles. If there are no
low-energy resonances in the fusion reaction, the astrophysical factor $S(E)$ 
is a slowly varying function of $E$ at low energies. For catalyzed fusion 
from the relative $s$-wave state of $N_1$ and $N_2$ the reaction factor  
$A(E)$ is defined as $A(E)=\sigma(E) v C_0^{-2}$ \cite{Jackson:1957zza}.  
Here $C_0^2$ is the $s$-wave Coulomb barrier penetration probability factor: 
\be
C_0^2=\frac{2\pi\eta_{12}}{e^{2\pi\eta_{12}}-1}\,.  
\label{eq:C02}
\ee
This gives 
\be
A(E)=\frac{S(E)}{\pi Z_1 Z_2 \alpha \mu c}\,(1-e^{-2\pi\eta_{12}})\,.
\vspace*{2mm}
\label{eq:A}
\ee
The transition from $S(E)$ to $A(E)$ takes into account the fact that 
the catalyst particle screens the Coulomb fields of the fusing nuclei 
and essentially eliminates the Coulomb barrier. 
The rates $\lambda$ of $X$CF reactions are related to the corresponding 
$A(E)$-factors as  
\be
\lambda=A(E)\rho_0\,,\qquad \rho_0\equiv \overline{|\psi_i(0)|^2_{R}}\,.
\label{eq:rates}
\ee
Here $\rho_0$ is the squared modulus of the 
atomic wave function of the initial ($N_1N_2X$) state taken at zero  
separation between the nuclei $N_1$ and $N_2$ 
(more precisely, at a distance of the order of the range of nuclear forces) and 
integrated over their distance $R$ to the $X$-particle. It plays the same 
role as the number density $n$ of the 
target particles in the usual expression for the reaction rates  
$\lambda=\sigma n v$ \cite{Jackson:1957zza}.

In muonic molecules or molecular ions, the energies of relative motion of 
nuclei are very low, so that $\eta_{12}\gg 1$; the term $e^{-2\pi\eta_{12}}$ 
in eq.~(\ref{eq:A}) is therefore always omitted in the literature on $\mu$CF. 
In addition, in evaluating the cross sections of the fusion reactions it is 
usually sufficient to consider the astrophysical $S$-factors in the limit 
$E\to 0$. 
   
In contrast to this, in $X$CF processes the kinetic energy $E$ of 
the relative motion of the nuclei $N_1$ and $N_2$ in ($XN_1N_2$) atoms  
is not negligible. 
For a system of charges bound by the Coulomb force, the virial theorem 
relates the mean kinetic energy $\bar{T}$ and mean potential
energy $\bar{U}$ as $2\bar{T}=-\bar{U}$. Therefore, the mean kinetic energy 
$\bar{T}$ in the ground state coincides with the binding energy of the 
system: $E_b=|\bar{T}+\bar{U}|=\bar{T}$. On the other hand, $\bar{T}$ is the 
sum of the mean kinetic energies of the relative motion of $N_1$ and $N_2$ 
and of their center-of-mass motion. In the $(ddX)$ atom, these two energies 
are to a good accuracy equal to each other, which gives $E\simeq E_b/2=0.145$ 
MeV. For other ($XN_1N_2)$ atomic systems of interest, the situation is more 
complicated; a rough estimate of the kinetic energy of relative motion 
of the fusing nuclei yields $E\simeq E_b m_2/(m_1+m_2)$, where $m_2$ is the 
smaller of the two masses.    
 
This can be explained as follows. Consider for simplicity the wave function 
of $(N_1N_2X)$ atom to be a product of the hydrogen-like wave functions of 
$(N_1X)$ and $(N_2X)$ systems, i.e.
\be
\Psi_i(\vec{r}_1, \vec{r}_2)=
\frac{1}{\pi(a_1a_2)^{3/2}} 
e^{-\frac{r_1}{a_1}-\frac{r_2}{a_2}}. 
\label{eq:psi_i}
\ee
Such a wave function obtains when one neglects the Coulomb interaction 
between $N_1$ and $N_2$, but in fact $\Psi_i$ of this form 
can partly include correlations between $N_1$ and $N_2$ provided that one 
replaces the charge of the $X$-particle $Z_X$ by an effective charge 
$Z_{X{\rm eff}}$ in the expressions for the Bohr radii $a_1$ and $a_2$ 
or, better still, treats $a_1$ and $a_2$ as variational parameters. 
The ground-state mean values of the total kinetic energy of the $(N_1N_2X)$ 
system, kinetic energy of the center of mass of $N_1$ and $N_2$ and  
their relative kinetic energy are then 
\begin{widetext}
\be
\bar{T}=
\frac{\hbar^2}{2m_1 a_1^2}+\frac{\hbar^2}{2m_2 a_2^2}\,,\qquad
\bar{T}_{\rm c.m.}=\frac{\hbar^2}{2m}\left(\frac{1}{a_1^2}+
\frac{1}{a_2^2}\right)\,,\qquad
\bar{T}_{\rm rel}\equiv E=\frac{\hbar^2}{2m_1 a_1^2}\frac{m_2}{m}+
\frac{\hbar^2}{2m_2 a_2^2}\frac{m_1}{m}\,, 
\label{eq:kinetic}
\ee 
\end{widetext}
where $m\equiv m_1+m_2$. Note that $\bar{T}_{\rm c.m.}+\bar{T}_{\rm rel}
=\bar{T}$. For $m_1=m_2$ we find $E=\bar{T}/2=E_b/2$. Assume now $m_1>m_2$. 
Since we expect $a_i\propto m_i^{-1}$ ($i=1,2$), the kinetic energies in 
eq.~(\ref{eq:kinetic}) can be estimated as 
\begin{widetext}
\be
\bar{T}\sim
\frac{\hbar^2}{2m_1 a_1^2}\,,\qquad\quad
\bar{T}_{\rm c.m.}\sim\frac{\hbar^2}{2ma_1^2}\,,\qquad\quad 
E= \bar{T}-\bar{T}_{\rm c.m.}\sim\, \frac{\hbar^2}{2m_1 a_1^2}\frac{m_2}{m}
\,\sim\, \bar{T}\frac{m_2}{m}\,, 
\label{eq:kinetic2}
\ee
\end{widetext}
which yields $E\sim E_bm_2/(m_1+m_2)$. 

Wave function (\ref{eq:psi_i}) can also be used for evaluating the parameter 
$\rho_0$ defined in eq.~(\ref{eq:rates}). Direct calculation gives 
\be
\rho_0=\frac{1}{\pi(a_1+a_2)^3}.
\label{eq:rho_0}
\ee
We shall also employ the wave function $\Psi_i(\vec{r}_1,\vec{r}_2)$ of 
eq.~(\ref{eq:psi_i}) in sec.~\ref{sec:IC} for evaluating of the parameter 
$\rho_1$ that enters into the expression for the rates of reactions 
(\ref{eq:r3c}) and (\ref{eq:r6c}).

\subsection{\label{sec:Sfact}Astrophysical $S$-factors and $X$CF reaction 
rates}

The Coulomb binding of the fusing nuclei $N_1$ and $N_2$ to an $X$-particle 
should have no effect on the strong interactions responsible for the fusion 
and can only modify the 
reaction rates due to the facts that (i) the Coulomb repulsion barrier is 
actually eliminated due to the very close distance between $N_1$ and $N_2$ in  
the $X$-atom, and (ii) because of the very small size of 
$X$-atoms, the number densities of $N_1$ and $N_2$ within an $(N_1N_2X)$-atom 
are many orders of magnitude larger than their number densities achievable 
for in-flight fusion. This means that, for those reactions that can occur in 
the absence of $X$-particles, we can use the experimentally measured values 
of the corresponding astrophysical $S$-factors in order to calculate the 
rates of the $X$CF reactions.  We take the relevant data from 
refs.~\cite{Pisanti:2020efz,Pisanti:2020efz,Angulo:1999zz,deSouza:2018gdx,
deSouza:2019pmr,Mukhamedzhanov:2011jr,Grassi:2017gud}. 
For internal conversion (IC) reactions (\ref{eq:r3c}) and 
(\ref{eq:r6c}), which do not have $X$-less analogues, we calculate the rates 
directly in the next subsection. The input data necessary for the calculations 
of the reaction factors $A(E)$ for all the discussed reactions and the 
obtained results are shown in Table~\ref{tab:rates1}.

\begin{table*}[t]
\centering
\begin{tabular}{|l|c|c|c|c|c|c|c|c|}
\hline
\hline
Reaction & $E_b$\,(MeV) & $E$\,(MeV) &  $S(E)$\,(MeV\,b) 
& $A(E)$\,(cm$^3$/s) & $\rho_{0,1}$ 
($10^{35}$\,cm$^{-3}$)
 & $\lambda$ (s$^{-1}$) \\
\hline
$(ddX)\to {\rm ^3He}+n+X$ & 0.290 
&  0.145  
& 0.102 \cite{Pisanti:2020efz} 
& $1.31\cdot 10^{-16}$  & 0.64 
 & $8.4\cdot 10^{18}$ \\
$(ddX)\to {\rm ^3H}+p+X$ & '' &  ''   
& $8.6\cdot 10^{-2}$ \cite{Pisanti:2020efz}  & 
$1.11\cdot 10^{-16}$    & '' & $7.1\cdot 10^{18}$ \\
$(ddX)\to {\rm ^4He}+\gamma+X $
& '' & ''  & $7\cdot 10^{-9}$ \cite{Angulo:1999zz} & 
~$9.0\cdot 10^{-24}$ & '' & $5.8\cdot 10^{11}$ \\
$(ddX)\to {\rm ^4He}+X$
& '' & ''  & 
$1.9\cdot 10^{-2}$
& 
$ 2.5\cdot 10^{-17}$  & 
0.73 & 
$1.8\cdot 10^{18}$ \\ 
$({\rm ^3He}\,dX)\to {\rm ^4He}+p+X$
& 0.91 
& 0.36 
& 7.1 \cite{deSouza:2018gdx}  & $4.0\cdot 10^{-15}$ 
& $7.6\cdot 10^{-3}$ 
& $3.0\cdot 10^{18}$ \\
$({\rm ^3H}dX)\to {\rm ^4He}+n+X $ 
& 0.32 & 0.13
& 5.6 \cite{deSouza:2019pmr}   & 
$6.2\cdot 10^{-15}$  & 
0.88 
& $5.5\cdot 10^{20}$ \\
$({\rm ^4He}\,dX)\to {\rm ^6Li}+\gamma+X $
& 1.16 & 0.39 
& $1.3\cdot 10^{-8}$ 
\cite{Grassi:2017gud}   
& ~$6.6\cdot 10^{-24}$ & 
$1.8\times 10^{-2}$ 
& $1.2\cdot 10^{10}$ \\
$
({\rm ^4He}\,dX)\to {\rm ^6Li}+X 
$ 
& '' & '' & 
1.55  
& $7.9\cdot 10^{-16}$   
& 0.14 
& $1.1\cdot 10^{19}$ \\
\hline
\end{tabular}
\caption{
Binding energies $E_b$, relative kinetic energies $E$ of 
fusing nuclei, astrophysical $S$-factors $S(E)$, reaction factors $A(E)$, 
$\rho$-parameters and rates $\lambda$ for the $X$CF 
reactions of stages I, II and III. Rates are inclusive of all sub-channels 
with either free or bound $X$-particles in the final state, except for 
IC processes, where final-state $X$ can only be free.  $\rho$-factors shown 
in the sixth column are values of $\rho_1$ for IC reactions (\ref{eq:r3c}) 
and (\ref{eq:r6c}) and $\rho_0$ for all other reactions. See text for details. 
}
\label{tab:rates1}
\end{table*}

\subsection{\label{sec:IC} Rates of internal conversion (IC) processes 
(\ref{eq:r3c}) and (\ref{eq:r6c})}  

IC is de-excitation of an excited nucleus in an atom through the ejection of 
an atomic electron (or, in the case of muonic atoms, of a muon) \cite{Rose}. 
For IC in a process of nuclear fusion, the initial excited 
nuclear state is a compound nucleus formed by the merger of the two fusing 
nuclei. In the case of $\mu$CF, the ejected particle is the muon; 
for $X$CF, it is more appropriate to speak about the ejection of the 
final-state nucleus itself rather than of the $X$-particle, as the latter is 
expected to be much heavier than light nuclei. However, when considered in 
terms of the relative motion between the ``nucleus'' and the orbiting 
particle, the treatment of IC in $X$CF closely parallels that in the case of 
the usual atoms or molecules.  

At low energies relevant to fusion of light nuclei, IC
predominantly proceeds through electric monopole (E0) 
transitions whenever this is allowed by angular momentum and parity 
selection rules. This is the case for reactions (\ref{eq:r3c}) and 
(\ref{eq:r6c}). 
The matrix element of an E0 transition can be written as \cite{akhBer}
\be 
M_{fi}= \frac{2\pi}{3}Z_X e^2 \tilde{Q}_0 
\psi_f^*(0)\psi_i(0)\,. 
\label{eq:Mfi} 
\ee 
Here $\psi_i(0)$ is the atomic wave function of the initial-state compound 
nucleus bound to the $X$-particle and $\psi_f(0)$ is the final-state 
continuum atomic wave function of the ejected nucleus, both taken at zero 
separation between the nucleus and the $X$ particle. 
The quantity $\tilde{Q}_0$ (not to be confused with $Q_0$ of 
table \ref{tab:general}) is the transition matrix element of the nuclear 
charge radius operator between the initial and final nuclear states: 
\be 
\tilde{Q}_0=\big\langle f\big|\sum_{i=1}^Z r_{pi}^2\big|i\big\rangle\,. 
\label{eq:tildeQ} 
\ee 
Here the sum is taken over nuclear protons. The rate of the 
process is readily found from the matrix element~(\ref{eq:Mfi}): 
\be 
\lambda_{\rm IC}=g_s\frac{8\pi}{9} Z_X^2 
\alpha^2\Big(\frac{mc}{\hbar}\Big)^2 c\, 
\sqrt{\frac{E_0}{2mc^2}}\;\big|\tilde{Q}_0\big|^2 F(Z_X Z, 
E_0)\,\rho_1\,. 
\vspace*{1.0mm} 
\label{eq:lambdaIC} 
\ee 
Here $g_s$ is the statistical weight factor depending on the angular momenta 
of the initial and final states, $m$ is the mass of the nucleus $N$ produced 
as a result of the fusion reaction and $E_0$ is its kinetic energy, which to 
a good accuracy coincides with the $Q$-value of the reaction. The factor 
$F(Z_X Z,E_0)$, defined as 
\be 
F(Z_X Z,E_0)=\frac{|\psi_f(0)|^2_Z~~~}{|\psi_f(0)|^2_{Z=0}}\,, 
\label{eq:F} 
\ee 
takes into account the deviation of the wave function 
of the final-state nucleus $N$ of charge $Z$ from the plane wave due to its 
interaction with the electric field of the $X$-particle. It is similar 
to the Fermi function employed in the theory of nuclear $\beta$-decay. 
By $\rho_1$ we denoted the quantity $|\psi_i(0)|^2$; it will be discussed in 
more detail below. 

We estimate the transition matrix elements of the charge radius operator as  
\be
\tilde{Q}_0\simeq r_i r_f\,,
\label{eq:tildeQ0}
\ee
where $r_f\equiv\langle r_c^2\rangle_f^{1/2}$ is the rms  charge radius 
of the final-state nucleus, and $r_i$ is the rms charge radius of the 
compound nucleus in the initial 
state, which we express through the rms charge radii $r_{N_{1c}}$ and 
$r_{N_{2c}}$ of the fusing nuclei $N_1$ and $N_2$ according to 
eq.~(\ref{eq:liqdr1}). Thus, we actually estimate the transition matrix 
element of the charge radius operator $\tilde{Q}_0$ as the geometric mean of 
the charge radii of the initial and final nuclear states. 

The function $F(Z_X Z, E_0)$ can be written as 
\be
F(Z_X Z, E_0)=\frac{2\pi\eta}{1-e^{-2\pi\eta}}\,. 
\label{eq:F2}
\ee
The Sommerfeld parameter $\eta$ is in this case 
\be
\eta=Z_X Z\alpha\,\sqrt{\frac{m c^2}{2E_0}}\,. 
\label{eq:eta2}
\ee
Note the similarity of eq.~(\ref{eq:F2}) with 
eq.~(\ref{eq:C02}); the difference is that eq.~(\ref{eq:C02}) describes the 
Coulomb repulsion of the like-sign charged nuclei $N_1$ and $N_2$, whereas 
eq.~(\ref{eq:F2}) accounts for the Coulomb attraction of the oppositely 
charged $X$-particle and final-state nucleus $N$ of charge $Z$. 
This attraction increases the value of the atomic wave function of $N$ at 
$r=0$ and leads to the enhancement of the reaction probability 
(Sommerfeld enhancement \cite{Sommerfeld}). 

The quantity $\rho_1$ in eq.~(\ref{eq:lambdaIC}) is the 
squared modulus of the wave function of the compound nucleus in the initial 
state of the IC reaction, taken at zero separation between the nucleus and the 
$X$-particle. We evaluate it by integrating the squared wave function of the 
initial atomic state $(N_1 N_2 X)$ over the distance $\vec{r}$ between the 
fusing nuclei in the volume $|\vec{r}|\le R_{1c}+R_{2c}$ (where $R_{1c}$ and 
$R_{2c}$ are the radii of the fusing nuclei) and setting the distance $R$ between 
their center of mass and the $X$-particle to zero:
\be
\rho_1\equiv \int
\limits
_{|\vec{r}|\le r_0}
|\psi_i(\vec{R}=0,\vec{r})|^2d^3r\,,\qquad 
r_0=R_{1c}+R_{2c}\,.
\label{eq:rho1-1}
\ee
For our estimates we use the wave function $\Psi_i(\vec{r}_1,\vec{r}_2)$  
defined in eq.~(\ref{eq:psi_i}). Going from the coordinates $\vec{r}_1$ and 
$\vec{r}_2$ of $N_1$ and $N_2$ to their relative coordinate $\vec{r}$ and 
c.m. coordinate $\vec{R}$ and substituting into (\ref{eq:rho1-1}), we obtain 
\begin{widetext}
\be 
\rho_1=\frac{a_*^3}{\pi(a_1a_2)^3}\Big[1-e^{-\frac{2r_0}{a_*}}
\big(1+2\frac{r_0}{a_*}+2\frac{r_0^2}{a_*^2}\big)\Big],\qquad\quad
a_*\equiv\left\{\frac{m_2}{m_1+m_2}a_1^{-1}+\frac{m_1}{m_1+m_2}
a_2^{-1}\right\}^{-1}\,.
\label{eq:rho1-2}
\ee
\end{widetext}

{\em Reaction $(ddX)\to {\rm ^4He}+X$ (\ref{eq:r3c}).} 
For this reaction $Z_1=Z_2=1$, $Z=2$, and $Q=23.56$ MeV.   
The transition matrix element of the charge radius operator, estimated 
according to eqs.~(\ref{eq:tildeQ0}) and (\ref{eq:liqdr1}), is 
$\tilde{Q}_0\simeq 4.52$ fm$^2$. For transitions from the ($ddX$) state one 
has to take into account that 
the initial state of two spin-1 deuterons in the atomic $s$-state 
can have total spin $S=2$ or 0 (spin 1 is excluded by Bose statistics). 
This gives 6 possible initial spin states. As the final-state nucleus 
$^4$He has zero spin and the transition operator is spin-independent, the 
IC transition (\ref{eq:r3c}) is only possible from the $S=0$ state of 
$(ddX)$. Therefore, $g_s=1/6$. For evaluating the parameter $\rho_1$ 
given in eq.~(\ref{eq:rho1-2}) we use the values of $a_1=a_2=8.53$\;fm found 
from the variational treatment of $(ddX)$ atom with wave function 
(\ref{eq:psi_i}). 

{\em Reaction 
$({\rm ^4He} d X)\to {\rm ^6Li}+X$ (\ref{eq:r6c})}. 
In this case $Z_1=2$, $Z_2=1$, $Z=3$, and $Q=0.320$ MeV. 
For the transition matrix element of the charge radius operator we find 
$\tilde{Q}_0\simeq 6.32$ fm$^2$.  
The reaction $({\rm ^4He}dX)\to {\rm ^6Li}+X$ is 
an E0 transition between nuclear states of total spin 1, therefore the 
weight factor $g_s=1$. For our evaluation of $\rho_1$ we take 
$a_1=1.81$\,fm, which is the Bohr radius of $({\rm ^4He}X)$ atom, and 
$a_2\simeq 30a_1$, as discussed in sec.~\ref{sec:posIons}. 

The expression for the rates of IC processes can conveniently be written in 
the form similar to (\ref{eq:rates}): $\lambda_{\rm IC}=A\rho_1$, 
where the reaction factor $A$ is defined as the factor multiplying $\rho_1$ 
in eq.~(\ref{eq:lambdaIC}). The values of the IC 
reaction factor $A$ and of the quantity $\rho_1$ for reactions (\ref{eq:r3c}) 
and (\ref{eq:r6c}) are presented in 
Table~\ref{tab:rates1}, along with the reaction factors and rates of the 
other discussed $X$CF reactions.  

To assess the accuracy of our calculations of the IC reaction factors, we 
compared our result for the $({\rm ^4He}dX)\to{\rm ^6Li}+X$ process with the 
existing calculations, which were carried out in the catalyzed BBN framework 
for the case of a singly charged catalyst particle $C$ using a simple scaling 
law \cite{Pospelov:2006sc} and within a sophisticated coupled-channel 
nuclear physics approach \cite{Hamaguchi:2007mp}. 
To this end, we re-calculated our result taking $Z_X=1$, $Q=1.3$ MeV and 
the c.m.\ energy $E=10$ keV which were used in 
\cite{Pospelov:2006sc,Hamaguchi:2007mp}. For the reaction factor $A$ of the 
process $d+({\rm ^4He} C)\to {\rm^6Li}+C$ we found 
$A\equiv \lambda_{\rm IC}/\rho_1\simeq 9.9\cdot 10^{-17}\;{\rm cm^3/s}$. 
Eq.~(\ref{eq:A}) then gives for the corresponding astrophysical $S$-factor 
$S(E)=0.19$ MeV\,b. This has to 
be compared with the results of ref.~\cite{Pospelov:2006sc} (0.3 MeV\,b) and 
ref.~\cite{Hamaguchi:2007mp} (0.043 MeV\,b). Our result lies between these 
two numbers, being a factor of 1.6 smaller than the former and a factor 
of 4.4 larger than the latter.

\begin{table*}[t]
\centering
\begin{tabular}{|l|c|c|c|c|c|c|}
\hline
\hline
Reaction & $Q_0$ (MeV) & $Q$ (MeV) & $\omega_{s0}$ & $\omega_s$ 
& $Br_0$ & $Br$ \\[0.2em]
\hline
$(ddX)\to {\rm^3He}+n+X$ & 
3.27 &  2.98  & $-$ & $-$ & 54.2\% 
& ~29.1\%   \\
$(ddX)\to ({\rm ^3He}X)+n$ & 
$-$ &  3.89  & 0.61 & 0.40 & $-$  
&~19.4\%   
 \\
$(ddX)\to {\rm^3H}+p+X$ &
4.03 &  3.74  & $-$ & $-$ & 45.8\% & 
34.4\%   \\
$(ddX)\to ({\rm^3H}X)+p$
& $-$ & 4.01 & 0.22 & 0.15 & $-$  
& ~6.2\%  \\    
$(ddX)\to {\rm^3H}+(pX)$
& $-$ & 3.84 & ~$1.9\cdot 10^{-2}$ 
& $1.2\cdot 10^{-2}$ & $-$ &   
~0.5\% \\
$(ddX)\to {\rm^4He}+\gamma+X $
& 23.85 
& 23.56 & $-$ & $-$  & 
$3.7\cdot 10^{-8}$ & 
~$4\cdot 10^{-9}$ \\
$(ddX)\to ({\rm^4He}X)+\gamma$
& $-$ & 24.71 & 
~0.95 & 0.87  & $-$ & 
$3\cdot 10^{-8}$ \\
$(ddX)\to {\rm^4He}+X$
& $-$ & 23.56 & $-$ & $-$ & $-$ 
& 10.4\% \\
$({\rm ^3He}\,dX)\to {\rm^4He}+p+X$
& 18.35 
& 
17.44 & $-$ & $-$ & 100\%  & 94\% \\
$({\rm ^3He}\,dX)\to ({\rm^4He}X)+p$
& $-$ & 18.60 & 0.29 & 0.06 & $-$ & 6\% \\
$({\rm ^3He}\,dX)\to {\rm ^4He}+(pX)$ & $-$ & 17.54 
& ~$2.3\cdot 10^{-3}$ & $3.0\cdot 10^{-4}$ & $-$ 
& $3.0\cdot 10^{-4}$ \\
$({\rm ^3H}dX)\to {\rm ^4He}+n+X $ & 17.59 & 17.27 
& $-$ & $-$ & 100\%  & 96\% \\
$({\rm ^3H}dX)\to ({\rm ^4He}X)+n$ 
& $-$ & 
18.42  & 0.23 & $ 4.0\cdot 10^{-2}$ & $-$ & 4.0\%  \\
$({\rm ^4He}\,dX)\to {\rm ^6Li}+\gamma+X $
& 1.475 & 0.32 & $-$ & $-$ & $ $100\%  & ~$10^{-13}$ \\
$({\rm ^4He}\,dX)\to ({\rm ^6Li}X)+\gamma $
& $-$ & 2.39
& $1-1.2\cdot 10^{-6}$ 
& ~$1-1.2\cdot 10^{-3}$  & $-$ & ~$1.9\cdot 10^{-8}$ 
\\
$
({\rm ^4He}\,dX)\to {\rm ^6Li}+X 
$
& $-$ & 0.32 & $-$ & $-$ & $-$ & $\simeq$100\% \\
\hline
\end{tabular}
\caption{
General characteristics of reactions (\ref{eq:r1a})-(\ref{eq:r6c}).  
$Q$ and $Br$ are $Q$-values and branching ratios of $X$CF reactions, $Q_0$ 
and $Br_0$ are values of these parameters for the corresponding 
$X$-less processes. $\omega_s$ and 
$\omega_{s0}$ are $X$-sticking probabilities found for nuclei of finite and 
zero radius, respectively.}
\label{tab:general}
\end{table*}

\subsection{Branching ratios}

The rates of the sub-channels of the $X$CF reactions 
in which the final-state $X$ sticks to one 
of the produced nuclear fragments are obtained by multiplying the total 
rate of the channel, given in Table~\ref{tab:rates1}, by the corresponding 
sticking probability $\omega_s$, shown in Table~\ref{tab:general}. 
The rate of the sub-channel with a free $X$ in the final state is then 
found by subtracting from the total rate of the channel the rates of all the 
sub-channels with bound $X$ in the final state. 
As an example, the rates of reactions 
(\ref{eq:r2b}) and (\ref{eq:r2c}) are obtained by multiplying the total rate 
of the $(ddX)$ fusion process with production of $^3$H and $p$,  
$\lambda=7.1\times 10^{18}$\,s$^{-1}$, by $\omega_s=0.15$ 
and $\omega_s=1.2\times 10^{-2}$, respectively; the rate of 
reaction (\ref{eq:r2a}) is then given by $\lambda(1-0.15-1.2\times 10^{-2})$. 
It is then straightforward to find the branching ratios of all the 
discussed reactions; the results are presented in Table~\ref{tab:general}.

\end{document}